\documentclass[aps,prd,preprint,a4paper,showpacs,showkeys,superscriptaddress]{revtex4-1}
\usepackage[latin9]{inputenc}
\usepackage{mathrsfs}
\usepackage{amsmath}
\usepackage{amssymb}
\usepackage{graphicx}
\usepackage{hyperref}
\hypersetup{colorlinks,%
	citecolor=blue,%
	linkcolor=cyan}
\begin{document}

\title{Investigations of strong cosmic censorship in 3-dimensional black strings}

\author{Jeongwon Ho}%
	\email[]{freejwho@gmail.com}%
	\affiliation{Center for Quantum Spacetime, Sogang University, Seoul 04107, Republic of Korea}%
\author{Wontae Kim}%
	\email[]{wtkim@sogang.ac.kr}%
       \affiliation{Center for Quantum Spacetime, Sogang University, Seoul 04107, Republic of Korea}%
	\affiliation{Department of Physics, Sogang University, Seoul, 04107,
		Republic of Korea}%
\author{Bum-Hoon Lee}%
	\email[]{bhl@sogang.ac.kr}%
      \affiliation{Center for Quantum Spacetime, Sogang University, Seoul 04107, Republic of Korea}%
	\affiliation{Department of Physics, Sogang University, Seoul, 04107,
		Republic of Korea}%


\begin{abstract}

Investigating the quasinormal modes of a massive scalar field on the 3-dimensional
black string (3dBS), we study the strong cosmic censorship (SCC) conjecture for the 3dBS in the T-dual relationship with the 3-dimensional rotating anti-de-Sitter (BTZ) black hole. It is shown that even though geometries of the two spacetimes are quite different, such as asymptotically AdS for the BTZ black hole and asymptotically flat for the 3dBS, the BTZ black hole and the 3dBS share similar properties for the SCC. Concretely speaking, the SCC conjecture can be violated even for asymptotically flat spacetime, i.e. the 3dBS.
These observations lead us to an assumption that the T-dual transformation preserves spacetime symmetries, at least, which are relevant to the SCC. In addition, we find a new feature of the quasinormal mode at the Cauchy horizon: in the case of the 3dBS, the spectral gap, 
$\alpha_{\mathrm{BS}}$ at the Cauchy horizon is not determined by the `{\it $\omega $-frequency mode}', but the `{\it m-frequency mode}'.

\end{abstract}

	
\keywords{}
	
\maketitle

\section{Introduction}

In general relativity, we believe that, given suitable initial data,
we can uniquely determine the geometry by solving Einstein's equations.
However, in the case of rotating or charged black holes, which have
an inner horizon serving as a boundary of future Cauchy development,
the story is not so simple.
Since the spacetime region beyond the Cauchy horizon (CH) is not uniquely determined by given initial data,
general relativity loses its predictive power.

About 40 years ago, Penrose \cite{penrose1968structure,simpson1973internal}
proposed a conjecture that perturbations incoming from outside the
event horizon are infinitely blueshifted at the CH and backreaction
makes the CH singular, so that beyond the CH the Einstein equation
ceases to make sense and the general relativity theory recovers its predictive power. 
It is called the strong cosmic censorship (SCC)
conjecture \cite{Penrose:1969pc,dewitt1974gravitational,Christodoulou:2008nj}.
Indeed, it has been shown that the SCC conjecture is viable for the
Reissner-Nordstr\"{o}m (RN) and Kerr black holes which are asymptotically flat
\cite{mcnamara1978instability,chandrasekhar1982crossing,Poisson:1989zz,Dafermos:2003wr,Hintz:2015koq,Franzen:2019fex}.

On the other hand, it has been known that perturbations outside the
event horizon decay with an inverse power law in an asymptotically
flat spacetime \cite{Price:1971fb,Nollert:1999ji,Kokkotas:1999bd,dafermos2016decay}.
Accordingly,  it
seems that the competition between the decay rate of perturbations
outside the event horizon and the amplification of blueshifted perturbation
determines the viability of the SCC conjecture.
In other words, these perturbations are infinitely blueshifted
at the CH, because in the case of the asymptotically flat spacetime background
they do not decay fast enough outside the event
horizon, so that the SCC is respected.

Such an observation that the validity of the SCC conjecture depends on the asymptotic geometry of black hole
naturally leads to studies on the SCC for the de-Sitter (dS) and the anti-de-Sitter (AdS) black holes
on which outside the event horizons perturbations exhibit an exponential decay \cite{Dyatlov:2010hq,Hintz:2016jak} and an inverse logarithmic decay \cite{Festuccia:2008zx,Holzegel:2011uu}, respectively.
Indeed, it has been shown that according to the decay rates of perturbations, the SCC conjecture is strengthened for RN-AdS \cite{Bhattacharjee:2016zof,Kehle:2018zws} and Kerr-AdS black holes \cite{Kehle:2020zfg} and is weakened for RN-dS black holes \cite{Cardoso:2017soq,Dias:2018etb,Luna:2019olw} (See also \cite{Chambers:1997ef,Brady:1998au}).
Specifically, the SCC conjecture is violated for the near extremal RN-dS black hole \cite{Cardoso:2017soq,Dias:2018etb,Luna:2019olw}.

From this, it can be seen that together with asymptotic geometries, the near extremal condition plays an essential role in determining the fate of the SCC conjecture. Such a result could be understood as follows: Taking the near extremal limit is effectively similar with applying the near horizon limit, which leads to an enhanced spacetime symmetry \cite{Balasubramanian:1998ee}. Since the enhanced symmetry gets the decay of perturbations faster outside the event horizon, the amplification of blueshifted perturbations would not be big enough for the perturbations to be singular at the CH. Thus, we may say that an enhancement in the spacetime symmetry has made the SCC conjecture fail.

From the same footing, it is expected that conversely, a reduction in the symmetry makes the SCC to be realized. Indeed, it has been shown that the photon sphere quasinormal mode, which cannot be found in the spherically symmetric RN-dS black hole background, exists in the axisymmetric Kerr-dS black hole background and decays sufficiently slowly to ensure that the SCC is respected for any nonextremal value of the black hole parameters \cite{Dias:2018ynt}.

This story continues with the BTZ black hole  \cite{Banados:1992wn}.
The logarithmic decay of perturbations is originated from the stable trapping phenomenon for 4-dimensional
asymptotically AdS black holes \cite{Holzegel:2011uu}. 
However, since the 3-dimensional general theory of relativity does not have gravitational dynamics,
the stable trapping phenomenon does not appear in the rotating BTZ black hole, 
which is a 3-dimensional AdS black hole. 
Since the factor causing the logarithmic decay of perturbations outside the event horizon disappears, it is expected that the SCC conjecture could be broken even though the BTZ black hole is asymptotically AdS. In fact,
Dias et. al. \cite{Dias:2019ery} have shown that the near extremal
rotating BTZ black hole badly violates the SCC conjecture (see also
\cite{Husain:1994xa,Chan:1994rs,Levi:2003cx,Bhattacharjee:2020gbo}).

In summary, it has been found that asymptotic geometries (signs of cosmological constants), rotations of black holes, near extremal limits for black hole parameters, and the number of spacetime dimensions play an important role in the SCC conjecture and they are more or less associated with the background spacetime symmetry. Thus, even though we do not have a unified description for the relation between the spacetime symmetry and the SCC, we can say that the background spacetime symmetry is essential for examining the SCC.

On the other hand, it is believed that in string theory, the T-dual transformation significantly changes string background geometries, but leaves unchanged the physics of the theory, i.e. all observable quantities in one description are identified with quantities in the dual description \cite{Bugden:2018pzv}. It is well known that in the context of the low energy string theory the 3-dimensional black string (3dBS) \cite{Horne:1991gn} is dual to the rotating BTZ black hole: a slight modification of the BTZ black hole solution yields an exact solution to the low energy string theory and then it is shown that under the T-dual transformation given in \cite{buscher1987symmetry} the rotating BTZ black hole is dual to the 3dBS \cite{horowitz1993string,Eghbali:2017ydo}.
It has been also shown that even though the BTZ black hole is asymptotically AdS and the 3dBS is asymptotically flat, their entropies are the same at least the leading and the next orders \cite{horowitz1994duality,Edelstein:2018ewc}.
 
In this paper, we study the viability of the SCC for the 3dBS using the quasinormal modes of a massive scalar field propagating on the 3dBS and compare this with the results of the rotating BTZ black hole. If the fact that the asymptotic geometry of 3dBS is flat is the most important factor associated with the SCC, the SCC will be respected as in the case of RN and Kerr black holes, and if the T-dual transformation preserves the spacetime symmetries, which are relevant to the SCC, it will not be viable at the near extremal limit, as in the case of the BTZ black hole.

Section 2 deals with reviewing the 3dBS solution and geometry. In Section 3, we calculate the quasinormal mode of a massive scalar field
on the 3dBS. The calculation process will proceed along the steps given in \cite{Dias:2019ery} for the BTZ black hole. The spectral
gap of the quasinormal mode is calculated and the SCC conjecture is examined in Section 4. Our discussions of the results obtained
in Section 4 and comments are given in Section 5.

\section{Three Dimensional Black Strings}

The low energy string action in three dimensions \cite{Horne:1991gn}
is given by
\begin{equation}
S=\int d^{3}x\sqrt{-g}e^{-2\phi}\left[\frac{4}{k}+R+4\left(\nabla\phi\right)^{2}-\frac{1}{12}H^{2}\right],\label{eq:loweea}
\end{equation}
where $\phi$ and $H$ are the dilaton and the three-form, respectively.
Since $H$ is closed, a two-form potential $B$ is defined by $H=dB$.
The variation of the action (\ref{eq:loweea}) with respect to the
metric, the antisymmetric field, and the dilaton gives
\begin{align}
 & R_{\mu\nu}+2\nabla_{\mu}\nabla_{\nu}\phi-\frac{1}{4}H_{\mu\lambda\sigma}H_{\nu}^{\lambda\sigma}=0,\nonumber \\
 & \nabla^{\mu}\left(e^{-2\phi}H_{\mu\nu\rho}\right)=0,\label{eq:eqofmloweea}\\
 & 4\nabla^{2}\phi-4\left(\nabla\phi\right)^{2}+\frac{4}{k}+R-\frac{1}{12}H^{2}=0.\nonumber
\end{align}
It has been shown \cite{horowitz1993string} that the BTZ black hole
with $\phi=0$ is a solution to the equations of motion in (\ref{eq:eqofmloweea}):
Substituting $\phi=0$, the second equation in (\ref{eq:eqofmloweea})
gives $H_{\mu\nu\rho}=(2/l)\epsilon_{\mu\nu\rho}$, where $l$ is
a constant with dimension of length and $\epsilon_{\mu\nu\rho}$ is
the volume form. Then, the first equation in (\ref{eq:eqofmloweea})
becomes
\begin{equation}
R_{\mu\nu}=-\frac{2}{l^{2}}g_{\mu\nu},\label{eq:3dEinsteineq}
\end{equation}
and from the third equation, we obtain $k=l^{2}$. Since the equation
(\ref{eq:3dEinsteineq}) is the 3-dimensional Einstein equation with
the negative cosmological constant $\Lambda=-1/l^{2}$, the BTZ black
hole
\begin{equation}
\begin{array}{cc}
  ds^{2}=\left(\mathcal{M}-\frac{\hat{r}^{2}}{l^{2}}\right)d\hat{t}^{2}-Jd\hat{t}d\varphi+\hat{r}^{2}d\varphi^{2}
 +\left(\frac{\hat{r}^{2}}{l^{2}}-\mathcal{M}+\frac{J^{2}}{4\hat{r}^{2}}\right)^{-1}d\hat{r}^{2}
\end{array}\label{eq:BTZ}
\end{equation}
is the solution to the equations (\ref{eq:eqofmloweea}). In equation
(\ref{eq:BTZ}), $\mathcal{M}$ is the mass and $J$ is the angular
momentum, which are related to the event horizon $\hat{r}_{+}$ and
the Cauchy horizon $\hat{r}_{-}$ given by
\begin{equation}
\mathcal{M}=\frac{\hat{r}_{+}^{2}+\hat{r}_{-}^{2}}{l^{2}},\quad J=\frac{2\hat{r}_{+}\hat{r}_{-}}{l}.
\end{equation}

Since the BTZ black hole solution ($\widetilde{g}_{\mu\nu}$, $\widetilde{B}_{\mu\nu}$,
$\widetilde{\phi}$) has the translational symmetry in the direction
of $\varphi$, another solution ($g_{\mu\nu}$, $B_{\mu\nu}$, $\phi$)
to the equation (\ref{eq:eqofmloweea}) can be obtained using the Abelian
T-dual transformation \cite{buscher1987symmetry}
\begin{align}
 & g_{\varphi\varphi}=\frac{1}{\widetilde{g}_{\varphi\varphi}},\:g_{\varphi a}=\frac{\widetilde{B}_{\varphi a}}{\widetilde{g}_{\varphi\varphi}},\nonumber \\
 & g_{ab}=\widetilde{g}_{ab}-\frac{\widetilde{g}_{\varphi a}\widetilde{g}_{\varphi b}-\widetilde{B}_{\varphi a}\widetilde{B}_{\varphi b}}{\widetilde{g}_{\varphi\varphi}},\nonumber \\
 & B_{\varphi a}=\frac{\widetilde{g}_{\varphi a}}{\widetilde{g}_{\varphi\varphi}},\:B_{ab}=\widetilde{B}_{ab}-\frac{2\widetilde{g}_{\varphi[a}\widetilde{B}_{b]\varphi}}{\widetilde{g}_{\varphi\varphi}},\nonumber \\
 & \phi=\widetilde{\phi}-\frac{1}{2}\ln\widetilde{g}_{\varphi\varphi},\label{eq:dualtransf}
\end{align}
where indices $a$ and $b$ denote $\hat{t}$ and $\hat{r}$. In \cite{horowitz1993string},
it has been shown that applying the transformation of (\ref{eq:dualtransf})
to the BTZ black hole (\ref{eq:BTZ}) and diagonalizing the metric
with new coordinates $t,\:x,\:r$ given by
\begin{equation}
\hat{t}=\frac{l\left(x-t\right)}{\left(\hat{r}_{+}^{2}-\hat{r}_{-}^{2}\right)^{1/2}},\;\varphi=\frac{\hat{r}_{+}^{2}t-\hat{r}_{-}^{2}x}{\left(\hat{r}_{+}^{2}-\hat{r}_{-}^{2}\right)^{1/2}},\:\hat{r}^{2}=lr,\label{eq:diagonalizing}
\end{equation}
one obtains the 3-dimensional black string solution
\begin{align}
ds^{2} & =-\left(1-\frac{M}{r}\right)dt^{2}+\left(1-\frac{Q^{2}}{Mr}\right)dx^{2}
  +\left(1-\frac{M}{r}\right)^{-1}\left(1-\frac{Q^{2}}{Mr}\right)^{-1}\frac{l^{2}dr^{2}}{4r^{2}},\label{eq:3dblackstring}\\
  B_{xt} & =Q/r,\quad\phi=-\ln\sqrt{rl},\label{eq:fields}
\end{align}
where $M$ and $Q$ are the mass and the axion charge per unit length
of the black string, respectively.

While the BTZ black hole is asymptotically AdS, the metric of the
black string (\ref{eq:3dblackstring}) is asymptotically flat. Such
geometric difference is originated from the dual transformation $\hat{r}^{2}\rightarrow1/\hat{r}^{2}$
in (\ref{eq:dualtransf}) and the AdS property has been transferred
to the dilaton field $e^{-2\phi}=rl$, which linearly increases with
$r$. Unlike differences in asymptotic geometry, the global causal
structure of the 3dBS in the non-extremal case of $0<\left|Q\right|<M$
is similar to that of the BTZ black hole. Not only the fact that
there are two horizons, but their positions are the same as for the
BTZ black hole, $r=r_{+}=M=\hat{r}_{+}^{2}/l$ and $r=r_{-}=Q^{2}/M=\hat{r}_{-}^{2}/l$.

\begin{figure}[h]
\begin{center}
\mbox{%
\includegraphics[scale=0.6]{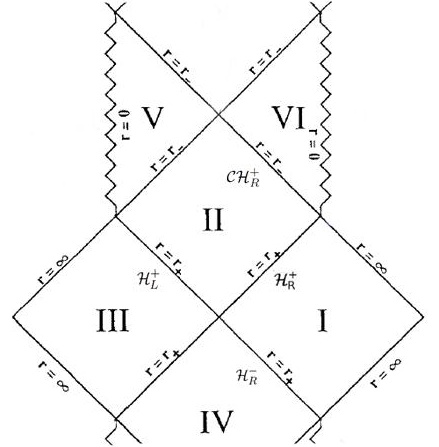}%
}\caption{\label{fig:PenD3dbs-1} Penrose diagram of the 3-dimensional black string:
(We ignore the region of $r<0$. See \cite{Horne:1991gn}.) Note
that a proper Penrose diagram of the black string cannot be drawn
on a flat surface like this figure.}
\end{center}
\end{figure}

The Penrose diagram of the 3dBS is given in Fig.~\ref{fig:PenD3dbs-1}.
In the Penrose diagram, a point represents a line in the $x$ direction in regions I - IV, while in regions V and VI, the line is along the $t$ direction. This means that we cannot draw a proper Penrose diagram of the 3dBS on a flat surface like Fig.~\ref{fig:PenD3dbs-1}. 

Such a global structure of the 3dBS shows that, unlike the case of black holes, the quasinormal mode cannot be written as analytical functions with only one set of Eddington-Finkelstein coordinates, which are regular at the event horizon and/or the CH. Instead, in order to represent the quasinormal mode analytic at the event horizon and the CH on the background of the 3dBS, two sets of Eddington-Finkelstein coordinates, which are regular at the event horizon and the CH, {\it respectively}, have to be introduced; Introducing the ingoing coordinate $v \equiv t+r^{*},$ where
\begin{equation}
r^{*}=\intop\left(1-\frac{M}{r}\right)^{-1}\left(1-\frac{Q^{2}}{Mr}\right)^{-1/2}\frac{l}{2r}dr,
\end{equation}
starting from region I, we can analytically extend the metric across the event horizon into region II. In the case of black holes, in order to take an analytical form of the quasinormal mode at the CH, in region II, converting the ingoing coordinate $v$ to the original time coordinate $t$, and converting again to the outgoing coordinate $u\equiv t-r^{*},$ which is regular at the CH \cite{Dias:2018ynt}. However, in the case of the 3dBS, in region VI, which is the region that we extend the metric from region II across the right CH, $t$ is not the time coordinate, but $x$ is. So, the outgoing coordinate $u$ is not regular at the CH in the 3dBS metric and we have to introduce another outgoing coordinate $\eta \equiv x-\bar{r},$ where
\begin{equation}
\bar{r}=\intop\left(\frac{M}{r}-1\right)^{-1/2}\left(1-\frac{Q^{2}}{Mr}\right)^{-1}\frac{l}{2r}dr.
\end{equation}
Approaching to the CH, the radial coordinate $\bar{r}$ behaves as
\begin{equation}
\bar{r}=\frac{1}{2\kappa_{-}}\ln\left|r-r_{-}\right|+constant,
\end{equation}
where $\kappa_{\pm}$ are surface gravities of the event horizon ($+$)
and the CH ($-$), respectively, given by
\begin{equation}
\kappa_{\pm}=\sqrt{\frac{r_{+}-r_{-}}{l^{2}r_{\pm}}}.\label{eq:surfacegravity}
\end{equation}

In summary, we introduce two sets of Eddington-Finkelstein coordinates $(u,~v),$ and $(\eta,\zeta)$, where
\begin{equation}
 u \equiv t - r^{*},~~v \equiv t+r^{*},~~ {\rm and},~~\eta \equiv x-\bar{r}, ~~ \zeta \equiv x + \bar{r}.
 \label{eq:EFcoord}
\end{equation}
Then, investigating regularity of the quasinormal mode near the CH, we will use these two sets of Eddington-Finkelstein coordinates and show that the spectral gap, which is essential to the SCC conjecture, is determined by the `{\it m-frequency mode}' associated with the outgoing coordinate $\eta $, not the `{\it $\omega $-frequency mode}' associated with the outgoing coordinate $u$.

\section{Quasinormal Modes of Massive Scalar Fields on 3dBS}

In this section, we will consider the behavior of quasinormal mode
of massive scalar fields at the CH. To do this, at first, we find
solutions of the Klein-Gordon equation for a massive scalar field
on the 3dBS. Then, a wavepacket is constructed with a linear combination
of these solutions, but its coefficients are selected to satisfy the
initial condition that the wavepacket is smooth at $\mathcal{H}_{\mathrm{L}}^{+}$,
$\mathcal{H}_{\mathrm{R}}^{-}$, and $\mathscr{I^{-}}$. The coefficients
$Z(\omega,m)$, $\mathcal{X}(\omega,m)$, and $\tilde{\chi}(\omega,m)$
contain the profile of such an initial condition of smoothness of
wavepackets on $\mathcal{H}_{\mathrm{L}}^{+}$, $\mathcal{H}_{\mathrm{R}}^{-}$,
and $\mathscr{I^{-}}$, respectively. Hereafter, we follow the notation
used in \cite{Dias:2019ery}. And the procedure of calculation in
this section is similar to the section 3 of \cite{Dias:2019ery}.
So it is recommended to refer to the detailed explanation provided
in this reference.

Consider a massive scalar field $\Phi=\Phi(t,x,r)$ with mass $\mu$
on the background fields ($g_{\mu\nu}$, $\phi$, $B$) given by (\ref{eq:3dblackstring})
and (\ref{eq:fields}). Given the matter action on this background
as \cite{Lee:1995vn,Kim:2001ev}
\begin{equation}
S_{\Phi}=\int d^{3}x\sqrt{-g}e^{-2\phi}\left[-\frac{1}{2}\left(\nabla\Phi\right)^{2}-\frac{\mu^{2}}{2}\Phi^{2}\right],\label{eq:matteraction}
\end{equation}
we obtain the Klein-Gordon equation from the action (\ref{eq:matteraction})
given by
\begin{equation}
\square\Phi-2\partial_{\mu}\phi\partial^{\mu}\Phi-\mu^{2}\Phi=0.\label{eq:kgequation}
\end{equation}

Here, we apply the T-dual transformation (\ref{eq:dualtransf}) only to the background fields ($g_{\mu\nu}$, $\phi$, $B$) and not to the scalar field $\Phi $, i.e., the massive scalar fields considered in \cite{Dias:2019ery} and in (\ref{eq:kgequation}) are not related with the T-dual transformation (\ref{eq:dualtransf}) each other.

Replacing the coordinate $r$ by the new radial coordinate $z=\frac{r-r_{-}}{r_{+}-r_{-}}$
and using $t$ and $x$-translational symmetries of the background
spacetime (\ref{eq:3dblackstring}), the scalar field can be decomposed
into the modes of frequencies $\omega$ and $m$ in these variables
\begin{align}
\Phi(t,x,z) & =e^{-i\omega t}e^{imx}R(z)\nonumber \\
 & =e^{-i\omega t}e^{imx}z^{-i\frac{m}{2\kappa_{-}}}(1-z)^{-i\frac{\omega}{2\kappa_{+}}}F(z).\label{eq:scalarfield}
\end{align}
Substituting the decomposed field (\ref{eq:scalarfield}) into the
Klein-Gordon equation (\ref{eq:kgequation}), we obtain the hypergeometric
function satisfied by $F(z)$ given by
\begin{equation}
z(1-z)F''(z)+[c-z(a+b+1)]F'(z)-ab\,F(z)=0,\label{eq:hypergeoeq}
\end{equation}
where
\begin{align}
a & =\frac{1}{2}\left(f(\omega,m)-i\frac{\omega}{\kappa_{+}}-i\frac{m}{\kappa_{-}}\right)\nonumber \\
b & =\frac{1}{2}\left(2-f(\omega,m)-i\frac{\omega}{\kappa_{+}}-i\frac{m}{\kappa_{-}}\right)\label{eq:constant}\\
c & =1-i\frac{m}{\kappa_{-}}\nonumber
\end{align}
and
\begin{equation}
f(\omega,m)\left(f(\omega,m)-2\right)=l^{2}\left(\mu^{2}+m^{2}-\omega^{2}\right)\label{eq:deltaeq}
\end{equation}

The hypergeometric equation (\ref{eq:hypergeoeq}) has three regular
singular points at $z=0,\;z=1,\;$and $z=\infty$, which correspond
to the CH, the event horizon, and the asymptotic infinity, respectively.
The general solution to the equation (\ref{eq:hypergeoeq}) can be
written in terms of a pair of linearly independent solutions about
each of these three regular singular points:

(i) In the vicinity of the CH, $z=0$,
\begin{align}
R_{{\rm out},-}= & z^{-\frac{1}{2}(1-c)}(1-z)^{\frac{1}{2}(a+b-c)}\,_{2}F_{1}(a,b;c;z),\nonumber \\
R_{{\rm in},-}= & z^{+\frac{1}{2}(1-c)}(1-z)^{\frac{1}{2}(a+b-c)}\,_{2}F_{1}(a-c+1,b-c+1;2-c;z).\label{eq:cauchybasis}
\end{align}

(ii) In the vicinity of the event horizon, $z=1$,
\begin{align}
R_{{\rm out},+}= & z^{-\frac{1}{2}(1-c)}|1-z|^{-\frac{1}{2}(a+b-c)} \, _{2}F_{1}(c-b,c-a;-a-b+c+1;1-z),\nonumber \\
R_{{\rm in},+}= & z^{-\frac{1}{2}(1-c)}|1-z|^{\frac{1}{2}(a+b-c)}\,_{2}F_{1}(a,b;a+b-c+1;1-z).\label{eq:cauchybasis_p}
\end{align}

(iii) At the asymptotic infinity, $z\rightarrow\infty$,
\begin{align}
R_{{\rm out},\infty}= & z^{-\frac{1}{2}(2a-c+1)}(z-1)^{\frac{1}{2}(a+b-c)}\,_{2}F_{1}\left(a,a-c+1;a-b+1;\frac{1}{z}\right),\nonumber \\
R_{{\rm in},\infty}= & z^{-\frac{1}{2}(2b-c+1)}(z-1)^{\frac{1}{2}(a+b-c)}\,_{2}F_{1}\left(b,b-c+1;-a+b+1;\frac{1}{z}\right). \label{eq:cauchybasis_i}
\end{align}

Each sets of basis solutions are related in the hypergeometric transformation
formula \cite{abramowitz1988handbook} as follows:
\begin{align}
 & R_{{\rm out},+}={\cal A}(\omega,m)R_{{\rm out},-}+{\cal B}(\omega,m)R_{{\rm in},-},\nonumber \\
 & R_{{\rm in},+}=\tilde{{\cal A}}(\omega,m)R_{{\rm in},-}+\tilde{{\cal B}}(\omega,m)R_{{\rm out},-},\nonumber \\
 & R_{{\rm in},+}=\frac{1}{{\cal T}(\omega,m)}R_{{\rm in},\infty}+\frac{{\cal R}(\omega,m)}{{\cal T}(\omega,m)}R_{{\rm out},\infty},\label{eq:transfrule}\\
 & R_{{\rm out},\infty}=\frac{1}{\tilde{{\cal T}}(\omega,m)}R_{{\rm out},+}+\frac{\tilde{{\cal R}}(\omega,m)}{\tilde{{\cal T}}(\omega,m)}R_{{\rm in},+},\nonumber
\end{align}
where the interaction coefficients ${\cal A}$, ${\cal B}$, ${\cal T}$,
${\cal R}$, and tildes of these coefficients are given by
\begin{align}
 & {\cal A}(\omega,m)=\frac{\Gamma(1-c)\Gamma(1-a-b+c)}{\Gamma(1-a)\Gamma(1-b)}\,,\nonumber \\
 & B(\omega,m)=\frac{\Gamma(c-1)\Gamma(-a-b+c+1)}{\Gamma(c-a)\Gamma(c-b)}\,,\nonumber \\
 & \tilde{{\cal A}}(\omega,m)=\frac{\Gamma(c-1)\Gamma(a+b-c+1)}{\Gamma(a)\Gamma(b)}\,,\label{eq:gamma1}\\
 & \tilde{{\cal B}}(\omega,m)=\frac{\Gamma(1-c)\Gamma(a+b-c+1)}{\Gamma(a-c+1)\Gamma(b-c+1)}\,,\nonumber
\end{align}
and
\begin{align}
 & {\cal T}(\omega,m)=\frac{\Gamma(a)\Gamma(a-c+1)}{\Gamma(a-b)\Gamma(a+b-c+1)}\,,\nonumber \\
 & {\cal R}(\omega,m)=\frac{\Gamma(a)\Gamma(b-a)\Gamma(a-c+1)}{\Gamma(b)\Gamma(a-b)\Gamma(b-c+1)}\,,\nonumber \\
 & \tilde{{\cal T}}(\omega,m)=\frac{\Gamma(a)\Gamma(a-c+1)}{\Gamma(a-b+1)\Gamma(a+b-c)}\,,\label{eq:gamma2}\\
 & \tilde{{\cal R}}(\omega,m)=\frac{\Gamma(a)\Gamma(a-c+1)\Gamma(-a-b+c)}{\Gamma(1-b)\Gamma(c-b)\Gamma(a+b-c)}\,.\nonumber
\end{align}

In order to investigate the singular behavior of the scalar field
at the CH, we need to find a unique solution satisfying the field
equation in regions I and II in the Penrose diagram in Fig.\ref{fig:PenD3dbs-1}.
First of all, we represent the wavepacket inside the event horizon
with the event horizon basis $(\Phi_{{\rm in},+},\Phi_{{\rm out},+})$
satisfying the initial data on $\mathcal{H}_{\mathrm{L}}^{+}$;
\begin{align}
\Phi(X) & \equiv\Phi_{{\rm out}}(X)+\Phi_{{\rm in}}(X)\nonumber \\
 & =\int\mathrm{d\omega d}m\,\left[Z(\omega,m)\Phi_{{\rm out},+}(\omega,m;X)+\tilde{Z}(\omega,m)\Phi_{{\rm in},+}(\omega,m;X)\right], \label{eq:insideeq}
\end{align}
where $X$ denotes spacetime coordinates. $\tilde{Z}(\omega,m)$ in
(\ref{eq:insideeq}) will be determined by the continuity condition
at the event horizon and will be represented with other coefficients,
which contain initial data on $\mathcal{H}_{\mathrm{R}}^{-}$ and $\mathscr{I^{-}}$.
From here onwards, we will omit variables of the field and the coefficients
such as $\Phi_{{\rm out},+}(\omega,m;X)\rightarrow\Phi_{{\rm out},+}$
and $Z(\omega,m)\rightarrow Z$. Using the transformation rule (\ref{eq:transfrule}),
we rewrite the solution (\ref{eq:insideeq}) in the CH basis $(\Phi_{{\rm in},-},\Phi_{{\rm out},-})$
\begin{align}
\Phi_{{\rm out}}(X)\equiv & \int d\omega\mathrm{d}m\left[Z{\cal A}+\tilde{Z}\tilde{{\cal B}}\right]\Phi_{{\rm out},-}\nonumber \\
\Phi_{{\rm in}}(X)\equiv & \int d\omega\mathrm{d}m\left[Z{\cal B}+\tilde{Z}\tilde{{\cal A}}\right]\Phi_{{\rm in},-}\label{eq:insideeq2}
\end{align}

On the other hand, a wavepacket outside the event horizon can be constructed
with the event horizon basis $(\Phi_{{\rm in},+},\Phi_{{\rm out},+})$
and the past null infinity basis $(\Phi_{{\rm in},\infty},\Phi_{{\rm out},\infty})$
with the initial data on $\mathcal{H}_{\mathrm{R}}^{-}$ and $\mathscr{I^{-}}$,
respectively,
\begin{align}
\Phi(X)= & \int d\omega\mathrm{d}m\,\left[\chi\left(\Phi_{{\rm out},+}+\tilde{{\cal R}}\Phi_{{\rm in},+}\right)+\tilde{\chi}\left(\Phi_{{\rm in},\infty}+{\cal R}\Phi_{{\rm out},\infty}\right)\right]
\label{eq:outsideeq}
\end{align}
Then, from the transformation rule (\ref{eq:transfrule}), the above
expression for the wavepacket (\ref{eq:outsideeq}) can be rewritten
as follows;
\begin{equation}
\Phi(X)=\int d\omega\mathrm{d}m\,\left[\chi\Phi_{{\rm out},+}+\left(\chi\tilde{{\cal R}}+\tilde{\chi}{\cal T}\right)\Phi_{{\rm in},+}\right].\label{eq:outsideeq2}
\end{equation}
Comparing (\ref{eq:outsideeq2}) with (\ref{eq:insideeq}), we obtain
the continuity condition at the event horizon $\mathcal{H}_{\mathrm{R}}^{+}$ given by
\begin{align}
\tilde{Z}(\omega,m)= & \tilde{\chi}(\omega,m){\cal T}(\omega,m)+\chi(\omega,m)\tilde{{\cal R}}(\omega,m)\label{eq:contcond}
\end{align}

\section{Strong Cosmic Censorship Conjecture in 3dBS}

In order to investigate the singular property of $\Phi(X)$
at the right Cauchy horizon $\mathcal{CH}_{\mathrm{R}}^{+}$, we rewrite the Cauchy horizon basis fields (\ref{eq:cauchybasis}) multiplied by mode frequency terms using the two outgoing coordinates $u$ and $\eta $ mentioned in Section 2 and in (\ref{eq:EFcoord}) up to the leading order given by
\begin{align}
\Phi_{{\rm out,-}}\approx & e^{-i\omega u}e^{im\eta}\left(1+\sqrt{z}\right)^{i\frac{\omega}{\kappa_{-}}}\mathcal{O}(1),\nonumber \\
\Phi_{{\rm in,-}}\approx & e^{-i\omega u}e^{im\eta}z^{+i\frac{m}{\kappa_{-}}}\left(1+\sqrt{z}\right)^{i\frac{\omega}{\kappa_{-}}}\mathcal{O}(1). \label{eq:fieldnearch}
\end{align}
Then, the wave packet $\Phi(X)$ is divided into in- and out-modes and is written as
\begin{align}
\Phi(X)= & \Phi_{{\rm out}}(X)+\Phi_{{\rm in}}(X),\nonumber \\
\Phi_{{\rm out}}(X)\approx & \int d\omega\int_{C_{\omega}}\mathrm{d}m\,{\cal G}_{\mathrm{out}}(\omega,m)e^{-i\omega u+im\eta}\mathcal{O}(1),\label{eq:packetnearch}\\
\Phi_{{\rm in}}(X)\approx & \int d\omega\int_{C_{\omega}}\mathrm{d}m\,{\cal G}_{\mathrm{in}}(\omega,m)e^{-i\omega u+im\eta}z^{+i\frac{m}{\kappa_{-}}}\mathcal{O}(1),\label{eq:packetnearchin}
\end{align}
where $C_{\omega}$ is the contour of integration with a fixed value
of $\omega$ and
\begin{align}
{\cal G}_{\mathrm{out}}(\omega,m)\equiv & Z{\cal A}+\left(\tilde{\chi}{\cal T}+\chi\tilde{{\cal R}}\right)\tilde{{\cal B}},\label{eq:modedensityout}\\
{\cal G}_{\mathrm{in}}(\omega,m)\equiv & Z{\cal B}+\left(\tilde{\chi}{\cal T}+\chi\tilde{{\cal R}}\right)\tilde{{\cal A}}.\label{eq:modedensityin}
\end{align}

As approaching to $\mathcal{CH}_{\mathrm{R}}^{+}$, \textit{i.e.}, $z\rightarrow0$,
$\Phi_{{\rm in,-}}$containing $z^{+i\frac{m}{\kappa_{-}}}$ becomes
singular, but $\Phi_{{\rm out,-}}$ is not. Thus, it is obvious that
any non-smooth behavior of $\Phi(X)$ at $\mathcal{CH}_{\mathrm{R}}^{+}$
arises from the in-mode $\Phi_{{\rm in}}$ of (\ref{eq:insideeq2}).
The smoothness of $\Phi_{{\rm in}}$ is determined by the longest
living quasinormal mode, \textit{i.e.}, the lowest $m$-frequency mode, which is associated with the outgoing coordinate $\eta $, not $u$.

Regarding the singular behavior of the quasinormal modes near the CH, it is valuable to check that which parts are changed by the T-dual transformation compared to the case of the BTZ black hole. Firstly, the fact that non-smooth behavior of normal modes occurs in the in-mode near the CH is the same in both cases of the BTZ and the 3dBS. Secondly, compared to the case of the BTZ black hole, in the field decomposition given in (\ref{eq:scalarfield}), the exponents of $z$ and $1-z$ have been changed with
\begin{align}
\frac{\omega - m \Omega_{-}}{\kappa_{-}} \rightarrow \frac{m}{\kappa_{-}}, ~~~~~
\frac{\omega - m \Omega_{+}}{\kappa_{+}} \rightarrow \frac{\omega}{\kappa_{+}},  \label{eq:cpbtz3dbs}
\end{align}
respectively. $\Omega_{\pm}$ are the angular velocities associated to each of the two horizons, respectively. As seen in (\ref{eq:fieldnearch}), in the case of the 3dBS, the in-mode $\Phi_{{\rm in,-}}$ can diverge near the CH, i..e., $z \sim 0$ because of the term  $z^{+i\frac{m}{\kappa_{-}}}$. Likewise, in the case of the BTZ black hole, the in-mode can diverge near the CH because of the term $z^{+i \frac{\omega - m \Omega_{-}}{\kappa_{-}}}$. This difference is directly related to the fact that in the case of the BTZ black hole,  the quasinormal mode propagating into the CH is the `{\it $\omega $-frequency mode}', but the `{\it m-frequency mode}' in the case of the 3dBS. That is, while in the case of the BTZ black hole the longest living quasinormal mode is determined by the lowest $\omega $-frequency mode for a given $m$, in the case of the 3dBS it is determined by the lowest $m$-frequency mode for a given $\omega $.

Now, we are ready to calculate the spectral gap $\alpha_{\mathrm{BS}}=-\mathrm{Im}(\mathit{m})$
using the longest living quasinormal mode.

Following the Christodoulou's argument for the SCC conjecture \cite{Christodoulou:2008nj}, we require the condition that the maximal Cauchy development is inextendible as a weak solution of the Einstein equation, a spacetime with locally square integrable Christoffel symbols, so that one cannot extend beyond the Cauchy horizon consistently with the equation of motion. For a linear scalar perturbation $\Phi $, the requirement is that $\Phi $ does not belong to the Sobolev space $H^1_{loc}$ at the CH. $H^1_{loc}$ is a space of functions to be locally square integrable, i.e. for any smooth compactly supported function $\psi $, if $(\Phi \psi)^2 + (\partial (\Phi \psi) )^2$ is integrable, then $\Phi \in  H^1_{loc}$  \cite{Luk:2015qja,Dafermos:2015bzz}. That is,
\begin{equation}
\int(r-r_{-})^{2\left(\frac{\alpha_{\mathrm{BS}}}{\kappa_{-}}-1\right)}dr
\end{equation}
diverges for
\begin{equation}
\beta\equiv\frac{\alpha_{\mathrm{BS}}}{\kappa_{-}}\leq\frac{1}{2}. \label{eq:christodoulouversionofscc}
\end{equation}
The spectral gap $\alpha_{\mathrm{BS}}$ is determined from the quantity ${\cal G}_{\mathrm{in}}(\omega,m)$
in (\ref{eq:modedensityin}).

\begin{figure}[h!]
\begin{center}
\mbox{%
\includegraphics[scale=0.3]{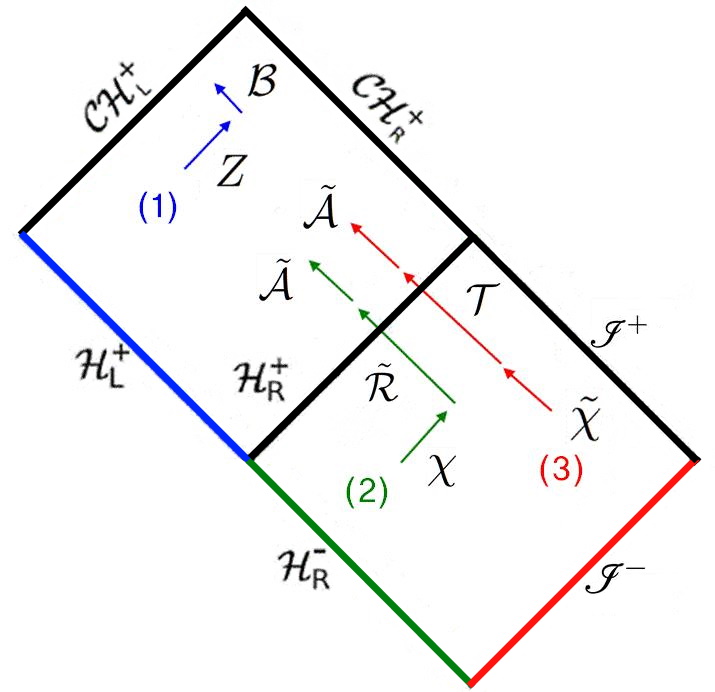}%
}\caption{\label{fig:InModesatCH} The blue wavepacket (1) comes out from $\mathcal{H}_{\mathrm{L}}^{+}$
with profile $Z$ and is reflected into $\mathcal{CH}_{\mathrm{R}}^{+}$.
The green wavepacket (2) comes out from $\mathcal{H}_{\mathrm{R}}^{-}$
with profile $\mathcal{X}$ and is reflected into $\mathcal{H}_{\mathrm{R}}^{+}$
and transmitted into $\mathcal{CH}_{\mathrm{R}}^{+}$. The red wavepacket
(3) comes in from $\mathscr{I^{-}}$ with profile $\tilde{\chi}$ and
is transmitted into $\mathcal{H}_{\mathrm{R}}^{+}$ and transmitted
into $\mathcal{CH}_{\mathrm{R}}^{+}$.}
\end{center}
\end{figure}

According to the initial condition that
the wavepacket is smooth at $\mathcal{H}_{\mathrm{L}}^{+}$, $\mathcal{H}_{\mathrm{R}}^{-}$,
and $\mathscr{I^{-}}$, and the continuity condition at the event horizon,
the profile functions $Z(\omega,m)$, $\mathcal{X}(\omega,m)$, and
$\tilde{\chi}(\omega,m)$ are smooth inside and outside of the event
horizon (see Fig.~\ref{fig:InModesatCH}). Thus, in (\ref{eq:modedensityin}),
we check singularities arising from ${\cal B},{\cal T}\tilde{{\cal A}}$ and
$\tilde{{\cal A}}\tilde{{\cal R}}$ in the lower half complex $m$-plane.

(A) From the reflection coefficient ${\cal B}$ and the transmission
coefficient $\tilde{{\cal A}}$, in common, poles at $m=0$ and $m=-in\kappa_{-}$
($n=0,1,2,\cdots$) appear,
\begin{equation}
\Gamma(c-1)=\Gamma\left(-i\frac{m}{\kappa_{-}}\right)=i\frac{\kappa_{-}}{m}\Gamma\left(1-i\frac{m}{\kappa_{-}}\right).
\end{equation}
Chosen the contour of integration $C_{\omega}$ in (\ref{eq:packetnearchin})
as passing below the pole at $m=0$, this pole does not give any contribution
deforming the contour to the lower half of the complex $m$-plane.
Since the contribution from the poles at $m=-in\kappa_{-}$ given
by the residue theorem is $z^{n}$ in (\ref{eq:packetnearchin}),
it vanishes smoothly at $\mathcal{CH}_{\mathrm{R}}^{+}$. Therefore,
the mode which comes out from $\mathcal{H}_{\mathrm{L}}^{+}$ and
is reflected to $\mathcal{CH}_{\mathrm{R}}^{+}$ does not give any
non-smooth property to $\Phi_{{\rm in}}(X)$.

(B) The transmission coefficient ${\cal T}$ and the reflection coefficient
$\tilde{{\cal R}}$ have the same poles at $m=m_{\mathrm{SD}}$ and
$m=m_{\mathrm{OD}}$ defined by
\begin{align}
\left[\frac{f(\omega,m)}{2}-\frac{i}{2}\left(\frac{\omega}{\kappa_{+}}-\frac{m}{\kappa_{-}}\right)\right]_{m=m_{\mathrm{SD}}}= & -n,\label{eq:sdmodeeq}\\
\left[\frac{f(\omega,m)}{2}-\frac{i}{2}\left(\frac{\omega}{\kappa_{+}}+\frac{m}{\kappa_{-}}\right)\right]_{m=m_{\mathrm{OD}}}= & -n,\label{eq:odmodeeq}
\end{align}
where $n=0,1,2,\cdots$. The subscripts SD and OD denote the `same
directional (SD) mode' and the `opposite directional (OD) mode', which correspond to the prograde mode and the retrograde mode in \cite{Dias:2019ery}, respectively.
From the equations (\ref{eq:sdmodeeq}) and (\ref{eq:odmodeeq}) with
(\ref{eq:deltaeq}), the slowest decaying frequencies of $m_{\mathrm{SD}}$
and $m_{\mathrm{OD}}$ of are obtained by
\begin{align}
\frac{m_{\mathrm{SD}}}{\kappa_{+}} & =\frac{\omega}{\kappa_{-}}+\sqrt{\rho}\sin\frac{\theta}{2}-i\left(\sqrt{\rho}\cos\frac{\theta}{2}-\frac{\kappa_{+}}{\kappa_{-}}\right),\label{eq:sdmodes}\\
\frac{m_{\mathrm{OD}}}{\kappa_{+}} & =-\frac{\omega}{\kappa_{-}}+\sqrt{\rho}\sin\frac{\theta}{2}-i\left(\sqrt{\rho}\cos\frac{\theta}{2}+\frac{\kappa_{+}}{\kappa_{-}}\right),\label{eq:odmodes}
\end{align}
where
\begin{align}
\rho= & \sqrt{\left(l^{2}\mu^{2}+\frac{r_{-}}{r_{+}}\right)^{2}+4l^{2}\left(1-\frac{r_{-}}{r_{+}}\right)\omega^{2}},\nonumber \\
\cos\frac{\theta}{2}= & \frac{1}{\sqrt{2}}\left\{ 1+\left(l^{2}\mu^{2}+\frac{r_{-}}{r_{+}}\right)\rho^{-1}\right\} ^{1/2}.\label{eq:defofrohandtheta}
\end{align}

For region I, the regularity condition at $\mathcal{H}_{\mathrm{R}}^{+}$
gives that $R_{{\rm in},+}$ is proportional to $R_{{\rm out},\infty}$
in (\ref{eq:transfrule}), then we find that the quasinormal mode
outside $\mathcal{H}_{\mathrm{R}}^{+}$ consists of SD and OD modes.
(Requiring the regularity condition at $\mathcal{H}_{\mathrm{R}}^{-}$,
there is another family of quasinormal modes consisting of the complex
conjugations of SD and OD modes.)

On the other hand, for region II, requiring the regularity condition
at $\mathcal{CH}_{\mathrm{R}}^{+}$, then $R_{{\rm in},+}$ is proportional
to $R_{{\rm out},-}$ in (\ref{eq:transfrule}). This gives in-out
quasinormal modes, which come in through $\mathcal{H}_{\mathrm{R}}^{+}$
and are reflected outward to $\mathcal{CH}_{\mathrm{R}}^{+}$, and
one of this family of quasinormal modes is the same as the OD mode
outside $\mathcal{H}_{\mathrm{R}}^{+}$. In addition, as the regularity
condition at $\mathcal{CH}_{\mathrm{L}}^{+}$, $R_{{\rm in},+}$ is
proportional to $R_{{\rm out},-}$ in (\ref{eq:transfrule}). This
gives in-in quasinormal modes, which come in through $\mathcal{H}_{\mathrm{R}}^{+}$
and are transmitted inward to $\mathcal{CH}_{\mathrm{L}}^{+}$, and
one of this family of quasinormal modes is the same as the SD mode
outside $\mathcal{H}_{\mathrm{R}}^{+}$.

A similar feature of the coincidence between the quasinormal modes inside and outside the event horizon has been shown in the rotating BTZ black hole as well \cite{Dias:2019ery}. However, the
matching structure of these modes is opposite: The SD (same-directional) mode in the 3dBS plays the same role as the retrograde (counter-rotating) mode in the BTZ, and the OD (opposite-directional) mode in the 3dBS plays the role of the prograde (co-rotating) mode in the BTZ. This is related to the fact that the quasinormal mode, which is regular at the CH of the 3dBS, is given by the `$m$-frequency mode', not by the `$\omega$-frequency mode'. 

According to such a correspondence, the divergence arising from the
pole at $m_{\mathrm{OD}}$ that appears from ${\cal T}$ and $\tilde{{\cal R}}$
is canceled with the zero of the transmission coefficient $\tilde{{\cal A}}$.
Thus, the spectral gap $\alpha_{\mathrm{BS}}$ is only determined
by the SD mode at $\mathcal{CH}_{\mathrm{R}}^{+}$. In addition, the imaginary part of $m_{\mathrm{SD}}$
depends on the frequency $\omega$. We consider the mode of $\omega=0$,
because the smoothness of the field is determined by the longest living
mode. All together the spectral gap is given by
\begin{equation}
\alpha_{\mathrm{BS}}=-\mathrm{Im}\left(m_{\mathrm{SD},\omega=0}\right)=\kappa_{+}\left(\sqrt{l^{2}\mu^{2}+\frac{r_{-}}{r_{+}}}-\sqrt{\frac{r_{-}}{r_{+}}}\right).\label{eq:spectralgap}
\end{equation}

Pushing down the contour of integration $C_{\omega}$ in (\ref{eq:packetnearchin})
up to a new contour $C$, which is a straight line in the lower half
complex $\omega$ plane of
\begin{equation}
\mathrm{Im}\left(m\right)=-\alpha_{\mathrm{BS}}-\epsilon
\end{equation}
where $\epsilon>0$, then picking up the contribution from the poles
that lie between $C_{\omega}$ and $C$, we obtain the contribution
from the pole at $m=-in\kappa_{-}$ that vanishes smoothly at $\mathcal{CH}_{\mathrm{R}}^{+}$
by $z^{n}$, as mentioned in (A), and the contribution from the pole
at $m_{(\mathrm{SD},\omega=0)}$ given by
\begin{align}
\Phi_{{\rm in}}(m_{(\mathrm{SD},\omega=0)})\approx & -2\pi i{\cal G}_{\mathrm{SD},\omega=0}\exp\left(im_{(\mathrm{SD},\omega=0)}\eta\right) z^{\beta+i\mathrm{Re}(m_{(\mathrm{SD},\omega=0)})/\kappa_{-}}\label{eq:inmodefinal}
\end{align}
where $\mathrm{Re}(m_{(\mathrm{SD},\omega=0)})$ is the real part of
$m_{(\mathrm{SD},\omega=0)}$, ${\cal G}_{\mathrm{SD},\omega=0}$ is
the residue of ${\cal G}_{\mathrm{in}}(\omega,m)$ at $m_{(\mathrm{SD},\omega=0)}$
and $\omega=0$, and
\begin{align}
\beta= & \frac{\alpha_{\mathrm{BS}}}{\kappa_{-}}=\sqrt{\frac{r_{-}}{r_{+}}}\left(\sqrt{l^{2}\mu^{2}+\frac{r_{-}}{r_{+}}}-\sqrt{\frac{r_{-}}{r_{+}}}\right).\label{eq:beta3dbs}
\end{align}

In \cite{Dias:2019ery}, for the rotating BTZ, they obtained $\beta$
given by
\begin{equation}
\beta=\frac{\triangle}{\frac{\hat{r}_{+}}{\hat{r}_{-}}-1}=\frac{\sqrt{l^{2}\mu^{2}+1}+1}{\frac{\hat{r}_{+}}{\hat{r}_{-}}-1}\label{eq:betabtz}
\end{equation}
where $\triangle$ is the conformal dimension of the operator dual
to the scalar field in the AdS/CFT correspondence and is determined
by the mass of the scalar field and the boundary condition. Here,
we apply the same scalar field mass $\mu$ for comparison. Taken the
extremal limit $r_{-}/r_{+}\rightarrow1$ for a given mass $\mu$, $\beta$ diverges in (\ref{eq:betabtz}). Thus, the
SCC is badly violated in this limit independent of the given value of $\mu$.

For the 3dBS in (\ref{eq:beta3dbs}), taking the extremal limit, $\beta $ becomes
\begin{equation}
\beta\sim\sqrt{l^{2}\mu^{2}+1}-1.\label{eq:betaextremallimit}
\end{equation}
Unlike with the case of rotating BTZ black holes, $\beta$ of the near extremal 3dBS in (\ref{eq:betaextremallimit}) depends
on the mass of the scalar field $\mu $. And because of this, even in the extremal limit, if the inequality
\begin{equation}
l^{2}\mu^{2}\leq\frac{5}{4}
\end{equation}
is satisfied, the SCC becomes viable for the 3dBS.
Of course, since still there is a big window $l^{2}\mu^{2}>5/4$ in which
the SCC could be violated near the extremal condition. Thus, the BTZ black hole and the 3dBS appear to share similar properties in the SCC.

On the other hand, under a far-from-extremal-condition, $r_{-}/r_{+}\ll1$
in (\ref{eq:beta3dbs}) and $\hat{r}_{-}/\hat{r}_{+}\ll1$ in (\ref{eq:betabtz}),
for a given mass $\mu$, $\beta$ approaches to zero, then the SCC
conjecture holds for the both the BTZ and the 3dBS.

Finally, in the equation (\ref{eq:beta3dbs}), taking the limit of $\mu \rightarrow 0$, $\beta $ becomes zero. Thus, for the massless scalar field, we see that there always exists the quasinormal mode, which lives long enough to diverge at the CH and realizes the SCC.

As a result, the T-dual transformation given by (\ref{eq:dualtransf}) that connects the BTZ black hole and the 3dBS generally preserves the spacetime symmetries related to the SCC, but it seems that the T-duality cannot preserve the physical properties of the scalar field propagating on these spacetime backgrounds. In fact, this result is obvious, because the T-duality considered here applies only to the background spacetimes, the BTZ black hole and the 3dBS, and not to the scalar fields on the background.

\section{Discussions}

By investigating the quasinormal mode of the massive scalar field on the 3dBS, we have studied the SCC conjecture for the 3dBS in the T-dual relationship with the rotating BTZ black hole. We have shown that the SCC is not viable in the extremal limit with $l^{2}\mu^{2}>5/4$ and is respected in the far-from-extremal condition. From this observation, we can say that even though geometries of the two spacetimes are quite different, the 3dBS and the BTZ black hole share the similar properties in the SCC and the T-dual transformation preserves spacetime symmetries, which could affect the SCC. It is another interesting result that the SCC conjecture can be violated even for an asymptotically flat spacetime, such as the 3dBS.

It is important to note that in this paper the T-dual transformation in (\ref{eq:dualtransf}) has been applied only to the background spacetimes, the BTZ black hole and the 3dBS, and not to the scalar fields on the background spacetimes. In other words, the scalar field on the 3dBS investigated in this paper is not directly related to the scalar field on the BTZ black hole considered in  \cite{Dias:2019ery} with the T-dual transformation. This leads to the difference in the feature of SCC between the BTZ black hole and the 3dBS as following; While in the case of rotating BTZ black holes, taking the extremal limit in (\ref{eq:betabtz}), one obtains $\beta \rightarrow \infty$ independent of the mass of the scalar field $\mu $, in the case of the 3dBS, $\beta $ has the mass dependence as can be seen in (\ref{eq:betaextremallimit}) in the extremal limit.

On the other hand, we have found another new feature of the quasinormal mode in the 3dBS: the spectral gap 
$\alpha_{\mathrm{BS}}$ (\ref{eq:spectralgap}) is determined in the $m$-complex plane not
in the $\omega$-complex plane. This is simply because the coordinate $\eta $ is regular at the CH and the spectral gap at the CH is determined by the `{\it m-frequency mode}' associated with the outgoing coordinate $\eta $, not the `{\it $\omega $-frequency mode}' associated with the outgoing coordinate $u$. And this feature is also related to the fact that the matching structure between the quasinormal modes inside and outside the event horizon is opposite to that in the rotating BTZ black hole.

Our study of the SCC for the 3dBS has been given for the classical massive scalar field. And we compared it with the case of the BTZ black hole in the T-dual relationship with the 3dBS in the classical level. It is likely that quantum perturbations reduce the spacetime symmetry and give an important effect to the SCC. Thus, it would be an interesting topic to investigate the preservation of the T-duality for the feature of the SCC at the quantum level, i.e. whether the effect of quantum perturbations in the 3dBS on the SCC is the same as that in the BTZ black hole \cite{Dias:2019ery,Hollands:2019whz,Emparan:2020rnp,Balasubramanian:2019qwk,Alishahiha:2021thv}. We leave this topic for a future work.

\vspace{0.4cm}


\noindent{\bf Acknowledgements}
\\
We are very grateful to Sang-Heon Yi for many helpful discussions.
Hospitality at APCTP during the program \textquotedblleft String theory, gravity
and cosmology (SGC2021)\textquotedblright{} is kindly acknowledged.
This research was supported by Basic Science Research Program through
the National Research Foundation of Korea(NRF) funded by the Korea
government(MSIP) (NRF-2017R1A2B2006159) and the Ministry of Education
through the Center for Quantum Spacetime (CQUeST) of Sogang University
(NRF-2020R1A6A1A03047877). JH was partially supported by NRF-2020R1A2C1014371. WTK was supported by the National Research Foundation of Korea(NRF) grant funded by the Korea government(MSIT) (No. NRF-2022R1A2C1002894)
BL was supported by the research program
NRF-2020R1F1A1075472.

\bibliographystyle{JHEP}

\newcommand{\noopsort}[1]{} \newcommand{\printfirst}[2]{#1}
  \newcommand{\singleletter}[1]{#1} \newcommand{\switchargs}[2]{#2#1}

\providecommand{\href}[2]{#2}\begingroup\raggedright\endgroup

\end{document}